\documentclass[twocolumn,showpacs,preprintnumbers,amsmath,amssymb]{revtex4-1}

\usepackage{graphicx}
\usepackage{dcolumn}
\usepackage{bm}
\usepackage{color}

\renewcommand{\bm}[1]{\mbox{\boldmath$#1$}}

\begin{document}

\preprint{APS/123-QED}

\title{Observation of multigap and coherence peak in the noncentrosymmetric superconductor CaPtAs: 
$^{75}$As nuclear quadrupole resonance measurement}

\author{Yuya Nagase$^1$, Masahiro Manago$^2$, Junichi Hayashi$^3$, Keiki Takeda$^3$, Hideki Tou$^1$, Eiichi Matsuoka$^1$, Hitoshi Sugawara$^1$, Hisatomo Harima$^1$, and Hisashi Kotegawa$^1$}

\affiliation{
$^1$Dept. of Phys., Kobe Univ., Kobe, Hyogo 657-8501, Japan \\
$^2$Dept. of Phys. and Mat., Shimane Univ., Matsue, Shimane 690-8504, Japan \\
$^3$Muroran Inst. of Tech., Muroran, Hokkaido 050-8585, Japan
}

\date{\today}

\begin{abstract}
We present synthesis and $^{75}$As-nuclear quadrupole resonance (NQR) measurements for the noncentrosymmetric superconductor CaPtAs with a superconducting transition temperature $T_c$ of $\sim 1.5$ K.
We discovered two different forms of CaPtAs during synthesis; one is a high-temperature tetragonal form that was previously reported, and the other is a low-temperature form consistent with the orthorhombic structure of CaPtP.
According to the $^{75}$As-NQR measurement for superconducting tetragonal CaPtAs, the nuclear spin-lattice relaxation rate $1/T_1$ has an obvious coherence peak below $T_c$ and does not follow a simple exponential variation at low temperatures. 
These findings indicate that CaPtAs is a multigap superconductor and a large $s$-wave component. 
\end{abstract}

\maketitle

In a noncentrosymmetric (NCS) metallic system, an antisymmetric spin-orbit interaction (ASOI) lifts the spin degeneracy of Fermi surfaces.
This lifting prevents an NCS superconductor from stabilizing the conventional spin-singlet or spin- triplet pairing when the band splitting is sufficiently larger than the superconducting (SC) gap, integrating both states in principle \cite{Gorkov,Frigeri,Samokhin,Fujimoto}.
A typical example of such an unusual SC state is heavy-fermion NCS superconductors such as CePt$_3$Si, CeRhSi$_3$, and CeIrSi$_3$ \cite{Bauer,Kimura,Sugitani}.
Because of the $f$-electron character, sufficient band splitting due to the ASOI is expected, and strong electronic correlations may assist to suppress the Cooper pair formation with simple $s$-wave symmetry.
Concretely, obvious indications supporting the unusual SC state, which differ from other centrosymmetric heavy-fermion superconductors, have been reported \cite{Kimura2,Yogi2004,Yogi2006,Mukuda}.

Except for $f$-electron systems, a significant ASOI is expected in $5d$ and $6p$ NCS systems. 
The perovskitelike cubic compound Li$_2$Pt$_3$B with an SC transition temperature $T_c$ of $\sim2.7$ K is a pioneering NCS superconductor in non-$f$-electron systems \cite{Badica}.
Penetration depth and nuclear magnetic resonance (NMR) measurements have suggested that a Cooper pair mainly exists in the spin-triplet state, and there exist line nodes in the gap function \cite{Yuan,Nishiyama}.
In contrast, the spin-singlet state is thought to be dominant in isostructural Li$_2$Pd$_3$B with a $T_c$ of $\sim7$ K, where a weak ASOI is expected \cite{Takeya,Yuan_pd,Harada}. 
In NMR, the coherence peak in the nuclear spin-lattice relaxation rate $1/T_1$ is strongly suppressed in Li$_2$Pt$_3$B but very visible in the Pd-system.
The suppression of the coherence peak suggests the presence of nodes, that is, the unconventional nature of the SC state.
Li$_2$Pt$_3$B is a unique non-$f$-electron system with no coherence peak, and such an SC state has been thought to be realized by the sufficient splitting of the spin degeneracy of Fermi surfaces.
Many NCS non-$f$-electron superconductors have been reported so far.
However, in most of them, the dominant $s$-wave state has been suggested from experiments including NMR \cite{Matano1,Matano2,Matano3}, showing difficulty in inducing the dominant triplet state in the absence of strong electronic correlations, as pointed out theoretically \cite{Samokhin2}.

Recently, interesting superconductivity has been discovered in NCS crystal CaPtAs, which undergoes an SC transition at $T_c=1.47$ K
\cite{Xie,Shang}. 
This material crystallizes in the tetragonal structure in the space group $I4_1md$ (the point group $C_{4v}$) \cite{Wenski}.
The band-structure calculation gives a band splitting of $E_{\rm ASOI} \sim 50-100$ meV because of the ASOI, which corresponds to $E_{\rm ASOI}/k_BT_c$ of $ \sim 400-800$.
This is smaller than that of CePt$_3$Si but comparable to that of Li$_2$Pt$_3$B \cite{Xie}.
The temperature dependence of superfluid density and specific heat in the SC state has been explained using a two-gap ($s+p$) model, indicating multigap nodal superconductivity with a dominant $p$-wave component \cite{Shang}.
The breaking of time-reversal symmetry as revealed by zero-field muon spin rotation ($\mu$SR) also suggests an unusual SC pairing in this material \cite{Shang}.
Measurements using other probes are also important to confirm the unusual SC state from a different viewpoint. 
Especially, the $1/T_1$ measurement in NMR, or nuclear quadrupole resonance (NQR), is a powerful tool to reveal the presence of $s$-wave character through the observation of a coherence peak, as demonstrated in the study of Li$_2$(Pd$_{1-x}$Pt$_x$)$_3$B \cite{Harada}.

In this paper, we present a method for synthesizing CaPtAs and an $^{75}$As-NQR study.
During synthesis, we discovered a different structural type of CaPtAs.
The known structure has a high-temperature tetragonal form and shows superconductivity ($\alpha$-CaPtAs), and the other has a low-temperature orthorhombic form, which is semiconducting ($\beta$-CaPtAs).
The $^{75}$As-NQR measurement using the high-temperature form has revealed that $T_1$ has an obvious coherence peak and exhibits a nonexponential temperature variation in the SC state.
These results suggest that $s$-wave symmetry is predominant and a multigap is formed in CaPtAs.

The sample used for the NQR measurement was prepared using a Bi-flux method and additional annealing.
The procedure for synthesis is described below.
Powder X-ray diffraction (XRD) was performed using an X-ray diffractometer with Cu K$\alpha$ radiation (Rigaku, MiniFlexII).
We also performed single-crystal XRD measurements using a Rigaku Saturn724 diffractometer with multilayer mirror monochromated Mo K$\alpha$ radiation at room temperature to evaluate the symmetry of crystals.
We measured electrical resistivity using a four-probe method, where the electrical contacts of wires were covered with silver paint.
The $^{75}$As NQR measurement (the nuclear spin $I=3/2$) was performed using powdered samples to obtain sufficient NQR intensity.
We used a $^3{\rm He}-^4$He dilution refrigerator for measurements at low temperatures, where the $\pi/2$ and $\pi$ pulses for the spin-echo method were applied for 13 $\mu$s and 25 $\mu$s, respectively. 
To avoid the heat-up effect in $T_1$ measurements, a sufficiently longer duration time than $T_1$ was taken between each pulse sequence.

\begin{figure}[htb]
\begin{center}
\includegraphics[width=\linewidth]{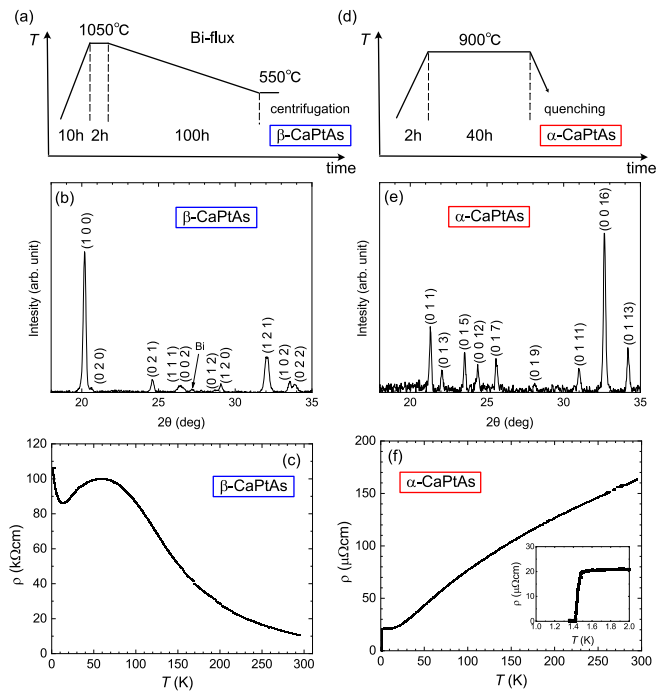}
\caption{Sequence of synthesis, XRD pattern, and temperature dependence of resistivity for different crystal structures of CaPtAs. (a-c) show the results for the low-temperature form $\beta$-CaPtAs, which is non-SC. (d-f) show the same set for the high-temperature form $\alpha$-CaPtAs. Centrifugation at 550$^{\circ}$C using the Bi-flux method yields $\beta$-CaPtAs. An annealing process at high temperatures is required to synthesize $\alpha$-CaPtAs. Centrifugation at high temperatures of approximately 800$^{\circ}$C is also effective to stabilize $\alpha$-CaPtAs.}
\label{xrd}
\end{center}
\end{figure}

\begin{table}[htb]
\begin{center}
\caption{Lattice parameters for $\beta$-CaPtAs (low-temperature form) and $\alpha$-CaPtAs (high-temperature form) determined the powder XRD measurements.}
\vspace{1ex}
\begin{tabular}{cc}\hline
$\beta$-CaPtAs & \\ \hline
crystal system & orthorhombic \\
space group & $Pmc2_1$, No.26 \\
$a$ (\AA) & 4.393(1) \\ 
$b$ (\AA) & 8.592(2) \\ 
$c$ (\AA) & 6.679(2) \\ \hline
\end{tabular}
\\
\vspace{2ex}
\begin{tabular}{cc}\hline
$\alpha$-CaPtAs & \\ \hline
crystal system & tetragonal \\
space group & $I4_1md$, No.109 \\
$a$ (\AA) & 4.1824(2) \\ 
$c$ (\AA) & 43.751(5) \\ \hline

\end{tabular}
\end{center}
\end{table}

The starting elements Ca, Pt, As, and Bi with a molar ratio of $1:1:1:5$ were put into an Al$_2$O$_3$ crucible and sealed in an evacuated quartz tube.
The tube was heated up to 1050$^{\circ}$C and then slowly cooled down to 550$^{\circ}$C at a rate of $-5^{\circ}$C/h.
After centrifugation, shiny platelike crystals were obtained.
This sequence is shown in Fig.~1(a). 
However, the powder XRD shown in Fig.~1(b) reveals that this crystal is not in $I4_1md$ but is consistent with the structure reported for orthorhombic CaPtP in the space group $Pmc2_1$ \cite{Wenski}.
The single-crystal XRD analysis also confirmed that this is in $Pmc2_1$.
The lattice parameters determined by powder XRD are presented in Table~I.
The cell volume is $\sim7$\% larger than that of CaPtP \cite{Wenski}. 
The sample contains crystals oriented in different directions and we could not find a single crystal that is sufficiently large for electrical resistivity measurement.
This material is also NCS, but it exhibits semiconducting behavior without superconductivity down to 1.3 K, as shown in Fig.~1(c).
As an additional procedure, we sealed the sample obtained through the first sequence in an evacuated quartz tube with a partial noble gas and annealed it at 900$^{\circ}$C for approximately 40 h, then quenched it with water.
This second sequence is shown in Fig.~1(d). 
The sample after this procedure crystallized in the tetragonal form of $I4_1md$ in an almost single phase, as shown in Fig.~1(e).
The lattice parameters in Table~I are consistent with the previous report \cite{Wenski}.
Our result reveals that CaPtAs has two different crystal structures.
It is thought that the sample obtained from the first sequence (second sequence) corresponds to a low-temperature phase (a high-temperature phase), respectively.
Hereafter, we call the high-temperature tetragonal form $\alpha$-CaPtAs and the low-temperature orthorhombic form $\beta$-CaPtAs.
The tetragonal $\alpha$-CaPtAs is also obtained directly by centrifugation at 800$^{\circ}$C in a similar process as the first sequence; therefore, a phase boundary is expected to be located between 550 and 800$^{\circ}$C.
As shown in Fig.~1(f), $\alpha$-CaPtAs exhibits metallic behavior and superconductivity below $T_c=1.47$ K, similar to a previous report \cite{Xie}.
Here, $T_c$ is defined as a 50\% decrease in the normal-state value.
The residual resistivity ratio (RRR) is estimated to be approximately 8, which is comparable to the previous one \cite{Xie}. 
We also tested the method reported in Ref.~22, where the sample was grown in a metallic crucible, but $\beta$-CaPtAs was typically produced, probably because the final cooling process was slow. 
Note that $\alpha$-CaPtAs is stable at room temperature and does not transform into the $\beta$ form for several months.
We performed a single-crystal XRD analysis on $\alpha$-CaPtAs, which is almost in a single phase, but this has not been well solved so far, probably because of a small amount of contamination caused by the low-temperature phase.
For the $^{75}$As NQR measurement, the sample was crushed into powder to obtain a sufficient NQR intensity in the SC state.

\begin{figure}[h]
\centering
\includegraphics[width=85mm,clip]{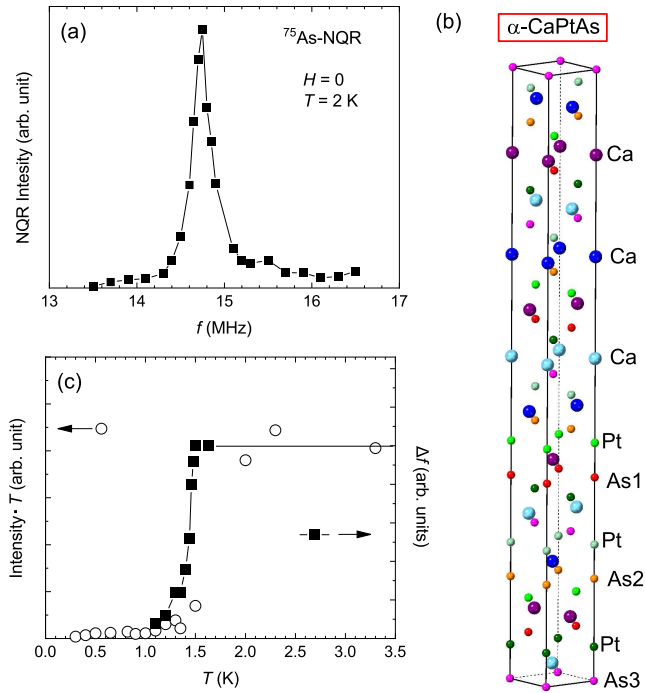}
\caption{(a) $^{75}$As-NQR spectrum of CaPtAs at 2 K. A signal of $\sim14.7$ MHz is considered the As3 site. (b) Crystal structure of tetragonal $\alpha$-CaPtAs. (c) Temperature dependence of the product of $^{75}$As NQR intensity and temperature, which is expected to be constant in the normal state (open circles). Diamagnetism measured by the NQR circuit is also shown (closed squares).}
\label{nqr}
\end{figure}

\begin{table}[b]
\caption{Calculated quadrupole frequency, $\nu_Q^{cal}$, asymmetry parameter $\eta^{cal}$, and resonance frequency $\nu_{res}^{cal}$ for the $\pm1/2 \leftrightarrow \pm3/2$ transition. Each As site is shown in Fig.~2(b).}
\vspace{1ex}
\begin{tabular}{ccccc}\hline
site & wyckoff & $\nu_Q^{cal}$ (MHz) & $\eta^{cal}$ & $\nu_{res}^{cal}$ (MHz) \\ \hline
As1 & 4a & 11.35 & 0.75 & 12.37 \\ 
As2 & 4a & 6.46 & 0.95 & 7.37 \\ 
As3 & 4a & 14.68 & 0.04 & 14.68 \\ \hline
\end{tabular}
\end{table}

\begin{figure}[b]
\begin{center}
\includegraphics[width=0.7\linewidth]{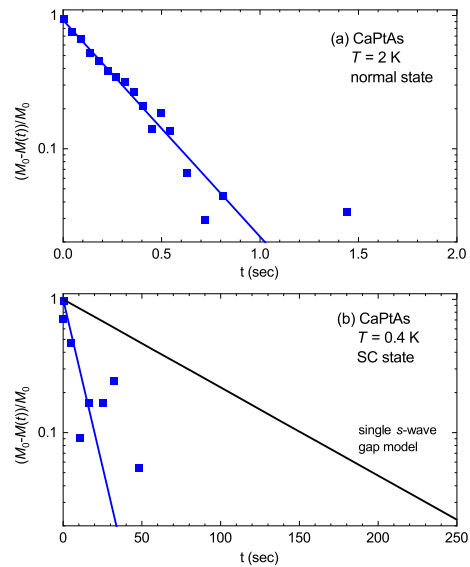}
\caption{The relaxation curves at (a) 2 K and (b) 0.4 K, measured for a signal at $\sim14.7$ MHz. The blue curves indicate the fitting results for the experimental data, using the single exponential function expected for the $\pm1/2 \leftrightarrow \pm3/2$ transition. In the SC state, the date obviously deviate from a curve expected for the $s$-wave model with a single isotropic gap.
}
\end{center}
\end{figure}

Figure 2(a) shows the frequency-swept $^{75}$As NQR spectrum of tetragonal $\alpha$-CaPtAs measured at 2 K.
As shown in Fig.~2(b), $\alpha$-CaPtAs has three As sites crystallographically, whose local symmetries are all [2mm.].
The calculated quadrupole frequency $\nu_Q^{cal}$, asymmetry parameter $\eta^{cal}$, and resonance frequency $\nu_{\rm res}^{cal}$ are obtained from a full-potential linear augmented plane wave (LAPW) calculation within the local density approximation (LDA), as shown in Table~II.
Here, we used the lattice parameters reported in Ref.~24.
The $\pm1/2 \leftrightarrow \pm3/2$ transition for an As site was observed at $\sim14.7$ MHz at zero fields, and this is consistent with the calculated $\nu_{\rm res}^{cal}$ for the As3 site.
The signals from other sites were observed in the NMR spectrum in a magnetic field (not shown).
We also checked the quadrupole frequencies for $\beta$-CaPtAs, which possesses two As sites, using our preliminary structural data.
They were both estimated to be more than 30 MHz, excluding a possibility that the signal at $\sim14.7$ MHz is partially composed of a different phase.
Figure 2(c) shows the temperature dependence of the NQR intensity multiplied by temperature, along with the SC diamagnetism detected using the NMR coil ($\Delta f$).
Diamagnetism starts to appear below $T_c$ of $\sim1.5$ K, and the intensity of the NQR signal at $\sim14.7$ MHz is strongly suppressed in the SC state because of the shielding of the rf pulse.
This ensures that the NQR signal at $\sim14.7$ MHz originates from the SC sample.
Note that this powdered sample has the same $T_c$ as the bulk sample.
There is no obvious disorder induced by the powdered sample.
If the SC pairing is unconventional for the NCS superconductor, it is expected that the disorder suppresses $T_c$ \cite{Mineev}.

The measurement of $T_1$ was conducted for a signal at $\sim14.7$ MHz.
The relaxation curves in the normal and the SC states are shown in Fig.~3.
In the normal state, the data follow the single exponential function well, ensuring the homogeneity of the electronic state of the present sample.
In the SC state, the data are somewhat scattered because of the weak signal, giving a relatively large error.

\begin{figure}[htb]
\begin{center}
\includegraphics[width=0.7\linewidth]{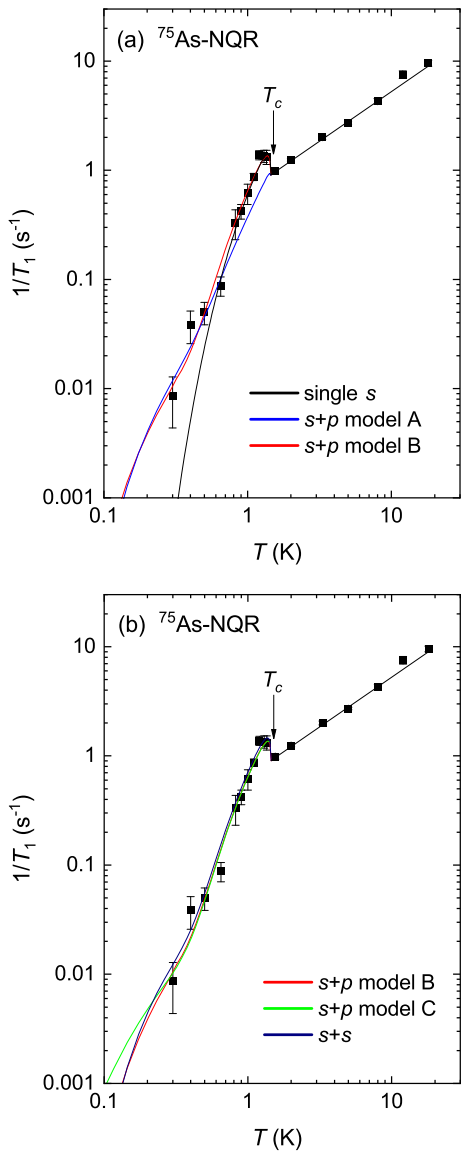}
\caption{(a)(b) Temperature dependence of $1/T_1$ for $^{75}$As-NQR. The straight line above $T_c$ indicates the Korringa $1/T_1\propto T$ relation, expected of typical metals. 
Each curve corresponds to the models shown in Table. III.}
\label{t1_point_line_ss}
\end{center}
\end{figure}

Figure 4 shows the temperature dependence of $1/T_1$ down to 300 mK. 
In the normal state above $T_c$, it exhibits the Korringa relation of $1/T_1T =const$ expected of typical metals.
This indicates that strong correlations in CaPtAs are absent.
The $1/T_1$ for the As nuclei, which has an electric quadrupole moment, can detect low-energy magnetic and electric fluctuations if they are present.
Superconductivity is thought to be mediated by a conventional electron-phonon interaction. 
In the SC state, a coherence peak is observed below $T_c$, which is a signature of $s$-wave symmetry.
The magnitude of the peak is close to twice the $1/T_1T$ value in the normal state, whose size is typically seen in conventional $s$-wave superconductors \cite{Masuda,Shimizu,Kawasaki}.
On the other hand, the data at low temperatures do not follow the typical $s$-wave model with a single isotropic gap, which is represented by the black curve.
The deviation is also clearly seen in the relaxation curve shown in Fig.~3(b).
This is consistent with the previous superfluid density and heat capacity measurements \cite{Shang}, supporting the multigap superconductivity of CaPtAs.

\begin{table}[htb]
\caption{Parameters for the SC gaps to examine the temperature dependence of $1/T_1$. $\Delta_0$ is the size of the SC gap, and the weight in the DOS for each band is shown. The major components are underlined.}
\label{t1_sp_point_tshang}
\vspace{1ex}
\begin{tabular}{cccc} \hline
\multicolumn{4}{l}{''single $s$''\ (black) } \\ \hline
symmetry \  & gap function \ & $\Delta_0/k_BT_c$ \ & weight \\ \hline
$\underline{s}$ & 1 & 2.0 & 1 \\ \hline
& & & \\ 
& & & \\ \hline
\multicolumn{4}{l}{''$s+p$: model A''\ (blue)} \\ \hline 
symmetry & gap function & $\Delta_0/k_BT_c$ & weight \\ \hline
$s$ & 1 & 0.4 & 0.15 \\ 
$\underline{p}$ & $\sin\theta$ & 1.97 & 0.85 \\ \hline
& & & \\ 
& & & \\ \hline
\multicolumn{4}{l}{''$s+p$: model B'' \ (red)} \\ \hline 
symmetry & gap function & $\Delta_0/k_BT_c$ & weight \\ \hline
$\underline{s}$ & 1 & 1.95 & 0.75 \\ 
$p$ & $\sin\theta$ & 0.4 & 0.25 \\ \hline
& & & \\ 
& & & \\ \hline
\multicolumn{4}{l}{''$s+p$: model C'' \ (green)} \\ \hline 
symmetry & gap function & $\Delta_0/k_BT_c$ & weight \\ \hline
$\underline{s}$ & 1 & 1.95 & 0.75 \\ 
$p$ & $\cos\theta$ & 0.4 & 0.25 \\ \hline
& & & \\ 
& & & \\ \hline
\multicolumn{4}{l}{$s^{++}$ \ (brown)} \\ \hline 
symmetry & gap function & $\Delta_0/k_BT_c$ & weight \\ \hline
$\underline{s}$ & 1 & 1.95 & 0.8 \\ 
$s$ & 1 & 0.4 & 0.2 \\ \hline
\end{tabular}

\end{table}

We evaluated several models to examine the SC symmetry from the result of $1/T_1$.
We also adopted a two-gap model similar to Ref.~23.
The blue curve in Fig.~3(a) is the $s$+$p$ (point node) model similar to that obtained in the specific heat (model A), where the $p$-wave is dominant \cite{Shang}.
The detailed parameters are listed in Table~III.
This could not reproduce the coherence peak because the weight of the density of the state (DOS) with an $s$-wave gap is significantly low.
The red curve is a model that reproduces our data well in the $s$+$p$ (point node) model (model B).
In this model, the majority band with a larger DOS is governed by $s$-wave symmetry to reproduce the coherence peak of the usual size.
SC gaps for $s$-wave and $p$-wave components are estimated to be $\Delta/k_BT_c = 1.95$ and $0.4$, respectively.
The sizes of the respective SC gaps are consistent with those estimated by specific heat \cite{Shang}; however, the SC symmetry of each band is the opposite.
It is important to determine the minor component of the SC symmetry; however, our data contain a considerable error well below $T_c$ because the NQR intensity in this temperature range is significantly low (Fig.~2(c)) and the long $T_1$ prevents the complete integration of the experimental data.
Therefore, there are constraints in determining the minor component of the gap function. 
As shown in Fig.~3(b), the $s$+$p$ (line node) model (model C) and $s$+$s$ model ($s^{++}$) also reproduce the temperature dependence of $1/T_1$ on the same level.
Note that the conventional $s^{++}$ symmetry should be excluded by observing the breaking of the time-reversal symmetry \cite{Shang}.

We mention a possibility that two bands originate in the band splitting due to the ASOI.
In this case, parity mixing can occur in each band and the order parameter $\Psi(\bm{k}) \pm |\bm{d}(\bm{k})|$ is expected. 
This has been proposed for CePt$_3$Si, whose point group $C_{4v}$ is the same as CaPtAs \cite{Frigeri2,Hayashi}.
If the $p$-wave component $\bm{d}(\bm{k})$ is dominant, a nodal gap and a nodeless gap are realized on respective bands and the coherence peak is weak or invisible \cite{Hayashi,Yanase}. 
This is inconsistent with our data. 
If the $s$-wave component $\Psi(\bm{k})$ is dominant, two nodeless gaps are realized and a difference in the gap sizes is caused by the $p$-wave component.
The coherence peak remains because of the large $s$-wave component.
This is compatible with our data.
The observation of the coherence peak contradicts the dominant $p$-wave state, but it does not deny the parity mixing.


In summary, we found that CaPtAs has two crystal structures and established a method for synthesizing each crystal selectively.
The high-temperature form $\alpha$-CaPtAs ($I4_1md$, No.109) can be fabricated using the Bi-flux method and by additional annealing treatment or centrifugation at high temperatures of approximately 800 $^{\circ}$C. 
The RRR of the present sample is approximately 8.
Improving the sample quality may not be easy because it is a metastable state at room temperature.
$\beta$-CaPtAs, which is in the low-temperature phase, exhibits semiconducting behavior without superconductivity down to 1.3 K.
In the $^{75}$As NQR measurement for SC $\alpha$-CaPtAs, we observed that $1/T_1$ in the normal state exhibits the Korringa relation, expected of typical metals.
In the SC state, $1/T_1$ showed a coherence peak below $T_c$, following nonexponential behavior at low temperatures. 
These two findings suggest that $\alpha$-CaPtAs is a multigap superconductor with a large $s$-wave component. 
Further investigation is desired, because the production of the multigap can be explained by the parity mixing with the $p$-wave component.

\section*{Acknowledgements}

The authors thank Y. Yanase for fruitful discussions.
This work was supported by JSPS KAKENHI Grant Number 18H04321 (J-Physics) and 21K03446.

\end{document}